\documentclass[preprint,12pt]{elsarticle}

\usepackage{amssymb}
\usepackage{epsfig}

\journal{Optics Communications}

\begin{document}

\begin{frontmatter}



\title{Storage and retrieval of time-entangled soliton trains in a three-level atom system coupled to an optical cavity}

\author[label1]{Davis D. M. Welakuh}
\ead{dwelakuh@yahoo.com}

\author[label1]{Alain M. Dikand\'e\corref{cor1}}
\cortext[cor1]{Corresponding author}
\ead{dikande.alain@ubuea.cm}

\address[label1]{Laboratory of Research on Advanced Materials and Nonlinear Science (LaRAMaNS), Department of Physics, Faculty of Science, University of Buea P.O Box 63 Buea, Cameroon}



\begin{abstract}
The storage and subsequent retrieval of coherent pulse trains in the quantum memory (i.e. cavity-dark state) of three-level $\Lambda$ atoms, are considered for an optical medium in which adiabatic photon transfer occurs under the condition of quantum impedance matching. The underlying mechanism is based on intracavity Electromagnetically-Induced Transparency, by which properties of a cavity filled with three-level $\Lambda$-type atoms are manipulated by an external control field. Under the impedance matching condition, we derive analytic expressions that suggest a complete transfer of an input field into the cavity-dark state by varying the mixing angle in a specific way, and its subsequent retrieval at a desired time. We illustrate the scheme by demonstrating the complete transfer and retrieval of a Gaussian, a single hyperbolic-secant and a periodic train of time-entangled hyperbolic-secant input photon pulses in the atom-cavity system. For the time-entangled hyperbolic-secant input field, a total controllability of the periodic evolution of the dark state population is made possible by changing the Rabi frequency of the classical driving field, thus allowing to alternately store and retrieve high-intensity photons from the optically dense Electromagnetically-Induced transparent medium. Such multiplexed photon states, which are expected to allow sharing quantum information among many users, are currently of very high demand for applications in long-distance and multiplexed quantum communication.
\end{abstract}


\end{frontmatter}


\section{Introduction}
\label{one}
Progress in the fabrication of materials with high intensity-dependent refractive index \cite{a1,a2,a3} has motivated a great deal of theoretical interest in nonlinear wave propagation in non-local media\cite{a4,a5}. In non-local optical systems such as thermal media \cite{a5}, the nonlinear optical response depends not only on the local intensity at a given point but also on the surrounding intensity profile. While many of these settings require high power laser lights, the phenomenon of electromagnetically-induced transparency (EIT) in multi-level atom systems \cite{a6} provides a cost-less and highly reliable mechanism to suppress photon loss, and to simultaneously increase light-matter interaction. Combined with sufficiently large nonlinearities, the phenomenon of EIT holds great potential for few-photon nonlinear optics and offers the possibility for many applications in communication and quantum information processing \cite{a6}.

EIT concretely is a quantum interference effect that permits the propagation of light through an otherwise opaque atomic medium. Interesting examples of quantum interference effects have been provided by recent reports on the observation of extremely slow group velocities \cite{a7}, as well as localization and containment of light pulse within an atomic cloud \cite{a8,a9}. In addition to a large suppression of optical absorption, EIT interference effect can be used to greatly enhance the efficiency of nonlinear optical processes. An addition to the noticeable effect of EIT is the significant change of dispersion property of optical media and the large reduction of the group velocity of optical wave packets \cite{a9,a10}. Slow light under conditions of EIT  has been observed in resonant multiple-level atoms, semiconductor quantum wells and quantum dots \cite{a11}. As a result of the enhancement of nonlinear optical processes, many remarkable phenomena such as giant Kerr nonlinearity \cite{a12,a13,a14,a15}, four-wave mixing, etc, have also been demonstrated \cite{a16,a17,a18}.

The most intriguing manifestation of nonlinearity in excited media is the existence of a specific kind of waves called solitons \cite{a19,a20}. These unique kind of wave packets occur because of a subtle balance of dispersion by nonlinearity, which results in their undistorted profiles during propagation over long distances. Optical solitons have been extensively investigated since they offer potential interesting applications in optical information processing and transmission, owing to their robustness during propagation. A large class of optical solitons can be produced with intense electromagnetic fields and via far-off resonant excitation schemes \cite{a21}. Optical solitons produced in such schemes travel with a velocity very close to the speed of light in vacuum.

In recent years the idea that slow-light solitons could emerge from three-level EIT systems has been proposed \cite{a22,a23}. With regards to quantum interference effects in EIT systems, considerable attention has been paid to the weak light solitons in atomic systems \cite{a24,a25}. For practical applications of optical memory it is desirable to obtain a probe pulse that is robust during its storage and retrieval. Previous analysis on storage of optical solitons has shown that a weak optical soliton pulse can be stored and retrieved in three-level atomic systems via a single EIT \cite{a26,a27,a28,a29,a30}, as well as in double EIT \cite{a31}. Investigations of the transfer of quantum correlations from traveling-wave light fields to collective atomic states have been carried out \cite{a32}. The transfer of single-photon quantum states to and from an optically dense coherently driven medium confined within a resonator, has been suggested \cite{a33}. In this last work it was demonstrated that well localized single-photon field, represented by the hyperbolic-secant pulse soliton, could be stored and retrieved by an adiabatic rotation of the cavity-dark state. The adiabatic transfer of the quantum state of photons to collective atomic excitations was brought into effect by intracavity EIT \cite{a34}, we note here that sources of single-photon wave packets have been proposed in ref. \cite{a35}.

In this work we shall be interested in the transfer, storage and retrieval of weak and high-intensity input photon wave packets in the quantum memory of a three-level $\Lambda$-atom system, under the condition of dynamical quantum impedance matching. We establish that the explicit dependance of analytic expressions of key parameters governing the transfer process, on the shape of the input wave packet, does not have negative effect on the dark state population but rather provides a well-defined normalization conditions for both the input and output wave packets. We suggest a new possible transfer scheme which involves a train of pulse solitons of finite period \cite{a36,a37,a38,a39} as the input field, instead of a single-pulse soliton \cite{a33}. That is, rather than propagating a single hyperbolic-secant pulse into the collective atom-cavity system as previously suggested \cite{a33}, we shall consider loading a continuous train of time-entangled high-intensity hyperbolic-secant pulses at some finite well-defined time period. By changing the Rabi frequency of the classical driving field the time-multiplexed input field  will be transferred into the cavity dark-state, and the population of the dark state is expected to change periodically with a kink profile over each period. The storage of multi-pulse photon solitons in three-level atoms has also been recently considered, using far-off-resonant Raman control scheme \cite{a40}. In our work we shall utilize the technique of adiabatic transfer \cite{a35} to map photonic states into collective atomic states. Worth mentioning, for this last scheme the non-destructive and reversible mapping of the quantum information contained in the photon pulses into collective atomic states is achieved using the technique of intracavity EIT \cite{a34}. Given that in the EIT technique properties of a cavity filled with three-level $\Lambda$-type atoms can be controlled by an external field, this enables the storage and retrieval of the periodic train of pulses by periodically switching off and on adiabatically the control field. Our results are particularly relevant for applications in quantum information processing involving relatively weak nonlinearity, where the phenomenon of modulational instability favors periodic optical solitons \cite{a38} even from a continuous-wave input field.

In section \ref{sectwo} we introduce the model and construct the families of dark states with one cavity photon from quantum states of the $\Lambda$-type atom system, coupled to an input field. We illustrate the consistency and reliability of the transfer scheme by considering two different low-intensity input fields, namely a Gaussian wave packet and plane waves. In section \ref{secthree} we propose a new dynamical quantum transfer scheme involving a periodic train of high-intensity input pulses, loaded in the three-level $\Lambda$ atom system at a finite and controllable time period. We formulate the probability-amplitude equations taking into account decays arising due to spontaneous emission, and investigate the time evolutions of characteristic quantities governing the storage and retrieval of the input soliton train. These include the mixing angle, the Rabi frequency and the dark state population. We discuss the connection of these characteristic parameters with results \cite{a33} for an hyperbolic-secant input pulse. Section \ref{secfour} will be devoted to concluding remarks.
\section{The $\Lambda$-type three-level atom system and dark state}
\label{sectwo}

The mechanism of adiabatic transfer and storage of photon states into a cavity-dark state and vice versa, has been described e.g. in refs. \cite{a6,a33}. Here we are interested in a distinct photon transfer mechanism which involves time-entangled high-intensity input photon solitons, our main objective being to extend the single-pulse soliton transfer scheme \cite{a33} to time-multiplexed periodic input photon pulses. We shall exploit the quantum impedance matching condition in order to obtain simplified relations for the transfer, storage and retrieval of any normalizad input field loaded into the atom-cavity system. 

\begin{figure}\centering
\includegraphics[width=3.5in, height=2.5in]{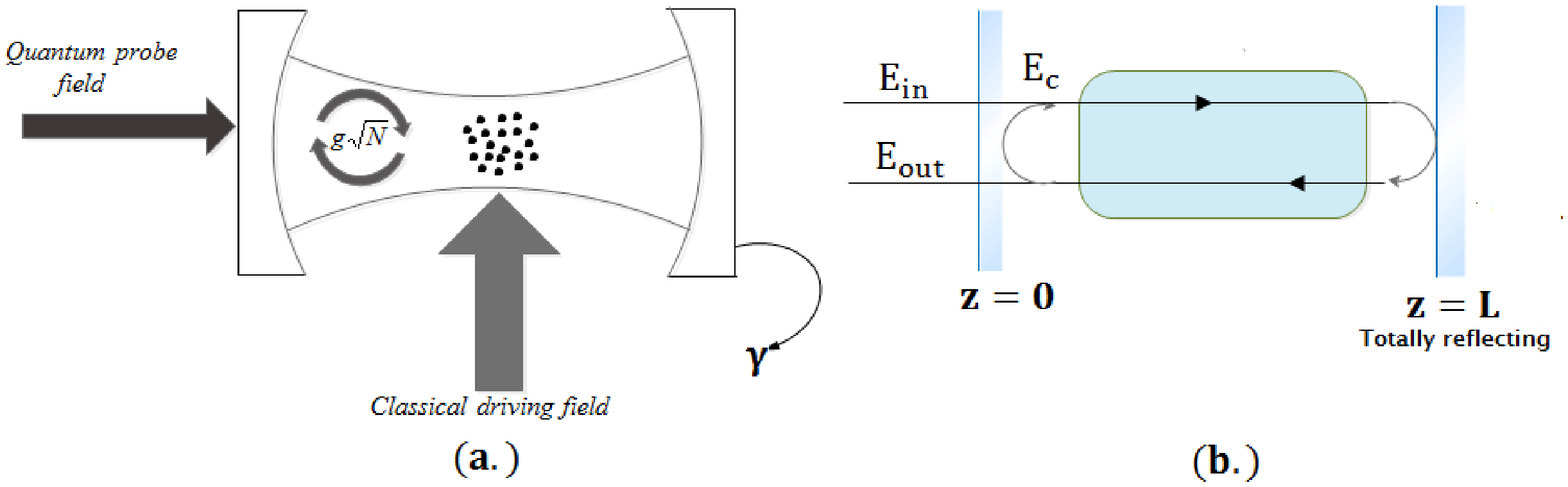}
\caption{\label{fig:one} (a.) Intracavity EIT setup with N atoms interacting with the cavity mode and with a classical driving field. $\gamma$ is the empty cavity decay rate. (b.) Simplified optical cavity setup with an optically dense ensemble of atoms. $E_{c}$, $E_{in}$, $E_{out}$ are the circulating, input, and output components of the field, respectively.}
\end{figure}

Consider an optically dense ensemble of N identical three-level atoms confined within an optical cavity. We assume an effective one-dimensional model consisting of a Fabry-Perot cavity with two mirrors, one partially transmitting the input field while the other mirror is totally reflecting, see fig.\ref{fig:one}. The input-output field is introduced as a continuum field modeled by a set of oscillator modes, denoted by the annihilation operator $\hat{b}_{k}$ ("$k$" free-field photon modes), coupled to the cavity mode with a coupling constant $\kappa$. The interaction of the cavity field $\hat{a}$ and the continuum of free-field modes $\hat{b}_{k}$ is described by the first term of Eq.(\ref{eq1}).

\begin{figure}\centering
\includegraphics[width=3.5in, height=2.5in]{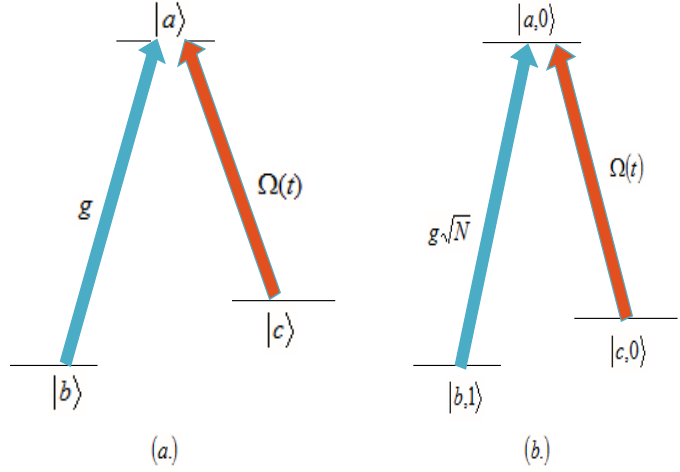}
\caption{\label{fig:two} (a.) The three-level atoms in $\Lambda$ configuration interacting with a quantized field and a classical field, with Rabi frequency $\Omega(t)$. The coupling constant between the quantum field and atoms is denoted by $g$. (b.) Singly excited mode of state $|b,1\rangle$ with $N$ three-level atoms in the basis of collective states.}
\end{figure}

Of the two dipole-allowed transitions one is coupled by a cavity mode with a coupling constant $g$, while the other optical transition is driven by a classical field in a coherent state with Rabi frequency $\Omega(t)$ (fig.\ref{fig:two}). 

The system is initially in state $|b\rangle$, and the fields cause transition between states. The dynamics of the coupled system is described by the Hamiltonian:
\begin{equation}
H = \hbar\kappa\sum_{k}\hat{a}^{\dagger}\hat{b}_{k} + \hbar g\sum_{i = 1}^N\hat{a}\sigma_{ab}^{i} + \hbar\Omega(t)e^{-i\nu t}\sum_{i = 1}^N\sigma_{ac}^{i} +  h.c., \label{eq1}
\end{equation}
where $\sigma_{\mu\nu}^{i}=|\mu\rangle_{ii}\langle\nu|$ is the flip operator of the $i^{th}$ atom between states $|\mu\rangle$ and $|\nu\rangle$ with $\mu,\nu=a,b,c$. $\hat{b}_{k}$ is the annihilation operator of a continuum of free-space modes of the single-photon field, coupled to the selected cavity mode by the creation operator $\hat{a}^{\dagger}$, while $\kappa$ describes the coupling of the selected modes. We define collective atomic operators \cite{a41} by the sum of flip operators;
\begin{equation}
\sigma_{ab}=\frac{1}{\sqrt{N}} \sum_{i=1}^{N}\sigma_{ab}^{i}, \quad \sigma_{ac}=\sum_{i=1}^{N}\sigma_{ac}^{i}, \label{eq2}
\end{equation}
with N the number of atoms. Note that these operators couple only symmetric Dicke-like states \cite{a33}. With the collective operators defined as above the interaction Hamiltonian describing the transitions between collective states (see fig.\ref{fig:two}) is given as:
\begin{equation}
H_{int} = \hbar g\sqrt{N}\hat{a}\sigma_{ab}+\hbar\Omega(t)e^{-i\nu t}\sigma_{ac} + h.c., \label{eq3}
\end{equation}
where the interaction strength of the cavity mode is enhanced by a factor of $\sqrt{N}$ which uplifts the stringent requirements of strong-coupling regime of cavity QED. Thus, single photons couple to collective excitations associated with EIT.  

Of special interest in the atom-cavity system is the dark state, which forms under the condition of two-photon resonance, when the energy difference between the metastable states equals the energy difference per photon of the two fields, i.e. when $\omega_{cb}=\nu-\nu_{c}$, with $\nu$ and $\nu_{c}$ the frequencies of the classical drive field and the cavity mode respectively. The resulting families of dark eigenstates with corresponding zero eigenvalues that contains one cavity photon is expressed as:
\begin{eqnarray}
|D,1\rangle &=& \frac{\Omega|b,1\rangle -  g\sqrt{N}|c,0\rangle}{\sqrt{\Omega^{2}+g^{2}N}} \nonumber \\
  &=& \cos\theta(t)|b,1\rangle - \sin\theta(t)|c,0\rangle, \label{eq4}
\end{eqnarray}
where $\cos\theta(t)=\frac{\Omega}{\sqrt {\Omega ^2  + g^2 N} }$ and $\sin\theta(t)=\frac{g\sqrt{N}}{\sqrt {\Omega ^2  + g^2 N} }$, with $\theta(t) = \arctan(\frac{g\sqrt{N}}{\Omega})$ the mixing angle. It is worth noting that in the limit $g\sqrt{N}\gg\Omega(t)$, the dark state $|D,1\rangle$ is nearly identical to $|c,0\rangle$ (i.e. $|D,1\rangle\sim|c,0\rangle$). In this limit, a single-photon excitation is shared among the atoms and the effective cavity-dark state decay is reduced as the dark state $|D,1\rangle$ contains only a very small component of $\left(\Omega/g\sqrt{N}\right)$ of the single-photon state $|b,1\rangle$ that is vulnerable to decay. 

We now discuss the principle of intracavity EIT relevant for our study. Three important mechanisms of dissipation and decay have to be distinguished. The dark state Eq.(\ref{eq4}) is immune against decay out of the excited states as it contains no component of the states, but is sensitive to decay of the coherence between the metastable states. This decay $\gamma_{bc}$ sets an upper limit for the lifetime of the dark state. In addition to this mechanism the effect of the finite lifetime of the cavity has to be considered. Thus, a bare cavity decay rate $\gamma$ leads to the effective decay rate $\gamma_{D}$ of the dark state $|D,1\rangle$ given by:
\begin{equation}
\gamma_{D} = \gamma\cos^{2}\theta(t).\label{eq10}
\end{equation}
This relation shows that varying the mixing angle $\theta(t)$ will influence the coupling of the cavity dark state to its environment. This is achieved in principle by changing the Rabi-frequency of the classical driving field $\Omega(t)$. This property of intracavity EIT will be used to effectively load the cavity system with an excitation resulting from incoming photon wave packets, and to subsequently release this energy after some storage time.

We now present a scheme of transfering a single-photon state of the input field into a single-photon cavity dark state (i.e. a single excitation of atom-cavity systems). We consider an input field in a general single-photon state i.e.:
\begin{equation}
|\Psi_{in}(t)\rangle = \sum_{k}\xi_{k}^{in}(t)\hat{b}_{k}^{\dagger}|0\rangle,\label{eq6}
\end{equation} 
where $\xi_{k}^{in}(t)=\xi_{k}^{in}(t_{0})e^{-i\omega_{k}(t-t_{0})}$, $|0\rangle$ is the vacuum state of the continuum of modes $\hat{b}_{k}$ and $\hat{b}_{k}^{\dagger}|0\rangle =|1_{k}\rangle$ is a bosonic Fock state which represents $|0,...,1_{k},...,0\rangle$, and $\sum_{k}|\xi_{k}^{in}|^{2}=1$. Hereafter we characterize these fields by an envelope wave function $\Phi_{in}(z,t)$ defined by:
\begin{equation}
\Phi_{in}(z,t) = \sum_{k}\langle 0_{k}|\hat{b}e^{ikz}|\Psi_{in}(t)\rangle = \frac{L}{2\pi c}\int d\omega_{k}\xi^{in}(\omega_{k},t)e^{ikz}.
\end{equation}
The normalization condition $(L/2\pi c)\int d\omega_{k}|\xi^{in}(\omega_{k},t)|^{2}=1$ of the Fourier coefficients implies the normalization of the input wave function according to:
\begin{equation}
\int\frac{dz}{L}|\Phi_{in}(z,t)|^{2}  = 1.
\end{equation}
Clearly, $\Phi_{in}(z,t)$ describes a single photon propagating along the $z$ axis.

The general state of the combined system of cavity mode and atoms, when the system interacts with the single-photon wave-packet, can be expressed in the compact form:
\begin{equation}
|\Psi(t)\rangle = \beta(t)|b,1,0_{k}\rangle + \alpha(t)|a,0,0_{k}\rangle + \epsilon(t)|c,0,0_{k}\rangle  
	+ \sum_{k}\xi_{k}(t)|b,0,1_{k}\rangle .\label{eq7}
\end{equation}
where $|b,1,0_{k}\rangle$ denotes the state corresponding to the atomic system in the collective state  $|b\rangle$, the cavity mode in single-photon state $|1\rangle$ and there are no photons in the outside modes $|0_{k}\rangle$. We stipulate two-photon resonance condition that requires the bare frequency of the cavity mode to coincide with the $a-b$ transition of the atoms, and the carrier frequency of the input wave packet i.e $\nu_{c}=\omega_{ab}\equiv\omega_{a}-\omega_{b}=\omega_{0}$ as well as the control field to be in resonance with the $a-c$ transition, i.e $\nu=\omega_{ac}$. The resulting equations of motion for the evolution of slowly-varying amplitudes in the rotating frame are:
\begin{eqnarray}
\dot{\alpha}(t) &=& -\frac{\gamma_{a}}{2}\alpha(t) - ig\sqrt{N}\beta(t) - i\Omega \epsilon(t), \label{eq11}\\
\dot{\beta}(t) &=& -ig\sqrt{N}\alpha(t) - i\kappa\sum_{k}\xi_{k}(t), \label{eq12}\\
\dot{\epsilon}(t) &=& -\frac{\gamma_{c}}{2}\epsilon(t) - i\Omega \alpha(t), \label{eq13}\\
\dot{\xi}_{k}(t) &=& -i\Delta_{k}\xi_{k}(t) - i\kappa \beta(t), \label{eq14} 
\end{eqnarray}
where $\Delta_{k}=\omega_{k}-\omega_{0}$ is the detuning of the free-field modes from the cavity resonance and $\omega_{0}=\omega_{ab}$. Here we have included a loss term $\gamma_{a}/2$ and $\gamma_{c}/2$ from state $|a\rangle$ and $|c\rangle$ respectively, representing spontaneous emission. In general, considering the motion of resonant atomic systems, the density matrix equations can be adopted. Nevertheless, for EIT-like coherent atomic systems the density matrix equations can be replaced by the probability amplitude equations without any difference \cite{a42}. Also Eq.(\ref{eq11}) is of no interest for it is not a constituent of the dark state, on the other hand Eqs.(\ref{eq12})-(\ref{eq14}) can be rewritten in terms of the dark and orthogonal bright states \cite{a45}:
\begin{eqnarray}
|D(t)\rangle &=& -i\cos\theta(t)|b,1,0_{k}\rangle + i\sin\theta(t)|c,0,0_{k}\rangle,\\
|B(t)\rangle &=& \sin\theta(t)|b,1,0_{k}\rangle + \cos\theta(t)|c,0,0_{k}\rangle,
\end{eqnarray}
which yields the following time-evolution equations for the dark and bright-state populations:
\begin{eqnarray}
\dot{D}(t) &=& -i\dot{\theta}(t)B(t) + \kappa\cos\theta(t)\sum_{k}\xi_{k}(t), \label{eq15}\\
\dot{\xi}_{k}(t) = &-& i\Delta_{k}\xi_{k}(t) - i\kappa\sin\theta(t)B(t) \kappa\cos\theta(t)D(t).\label{eq16}
\end{eqnarray}
Following the adiabatic elimination of the bright state and non-adiabatic corrections \cite{a33}, the remaining amplitudes of dark states and free-field components are given by:
\begin{eqnarray}
\dot{D}(t) &=& \kappa\cos\theta(t)\sum_{k}\xi_{k}(t),\label{eq17}\\
\dot{\xi}_{k}(t) &=& -i\Delta_{k}\xi_{k}(t) - \kappa\cos\theta(t)D(t).\label{eq18}
\end{eqnarray}
By formally integrating Eq.(\ref{eq17}) in the continuum limit, substituting the result into Eq.(\ref{eq18}) and invoking the standard Markov approximation assuming that no photon arrives the cavity before some reference time $t_{0}$, the dark state and output field are found as:
\begin{equation}
D(t) = \sqrt{\gamma\frac{c}{L}}\int_{t_0}^{t}d\tau\,\cos\theta(\tau)\Phi_{in}(0,\tau)
e^{-\left[\frac{\gamma}{2}\int_{\tau}^{t}\cos^2\theta(\tau')d\tau'\right]}, 
\label{eq19} 
\end{equation}
\begin{eqnarray}
\Phi_{out}(0,t) &=& \Phi_{in}(0,t) - G(t), \label{eq20}
\end{eqnarray}
\begin{equation}
G(t) = \gamma\cos\theta(t)\int_{t_{0}}^{t}d\tau\cos\theta(\tau)\Phi_{in}(0,\tau) e^{-\left[\frac{\gamma}{2}\int_{\tau}^{t}\cos^2\theta(\tau')d\tau'\right]}, \label{numa}
\end{equation}
where $\gamma=\frac{\kappa^{2}L}{c}$ is the empty cavity decay rate. In order to have complete transfer of free-field photons into the dark state, we require an optimization of $\cos\theta(t)$ such that from Eq.(\ref{eq19}), $D(t) \sim \int_{t_{0}}^{t}\Phi_{in}(0,\tau)d\tau$. It will be shown later that a result of optimization of the time-dependence of $\cos\theta(t)$ yields the condition for the normalization of the input field $\Phi_{in}(t)$, which ensures that the dark-state population tends to unity for each input field. To capture and subsequently release a single-photon state of the light field in this way, we start by accumulating the field in a cavity mode and then adiabatically switching off the driving field in such a way that an initial free-space wave packet can be stored in a long-lived atom-like dark state. By adiabatically switching on the Rabi-frequency of the classical driving field, we can release the stored wave packet. 

The optimization of $\cos\theta(t)$ in Eq.(\ref{eq19}) is acheived under conditions of quantum impedance matching. Taking advantage of the destructive interference of the directly reflected and circulating field components within the cavity, we shall require $\Phi_{out}=\dot{\Phi}_{out}=0$ which leads to:
\begin{equation}
-\frac{d}{dt}\ln\cos\theta(t) + \frac{d}{dt}\ln\Phi_{in}(t) = \frac{\gamma}{2}\cos^{2}\theta(t). \label{eq21}
\end{equation}
The above equation is referred to as quantum or dynamical impedance matching. The first term on the left hand side of Eq.(\ref{eq21}) describes internal losses due to coherent Raman adiabatic passage, while the second term appears due to time dependence of the input field $\Phi_{in}$. The right hand side of Eq.(\ref{eq21}) can be interpreted as an effective cavity decay rate which is reduced due to intracavity EIT. Solving Eq.(\ref{eq21}) for $\cos\theta(t)$ leads to:
\begin{equation}
\cos\theta(t) = \frac{1}{\sqrt{\gamma }}\frac{\Phi_{in}(t)}{\sqrt{\int_{t_{0}}^{t}\Phi_{in}^{2}(t^{'})dt^{'}}}.
\label{eq22}
\end{equation}
Quite remarkably, impedance matching presents a viable technique for complete transfer of single-photon state of the free-field into the cavity dark-state by optimizing $\cos\theta(t)$ as in Eq.(\ref{eq22}). The corresponding optimization of $\cos\theta(t)$ is in principle achieved by changing the Rabi frequency of the classical driving field, which in essence is equivalent to varying the mixing angle $\theta(t)$. The Rabi frequency of the classical driving field that optimizes the time-dependence of $\cos\theta(t)$ derived from Eq.(\ref{eq22}) is given by:
\begin{equation}
\Omega(t) = g\sqrt{N} \frac{\Phi_{in}(t)}{\sqrt{\gamma\int_{t_{0}}^{t}\Phi_{in}^{2}(t^{'})dt^{'} - \Phi_{in}^{2}(t)}}. \label{eq23}
\end{equation}
With this choice of the driving field, the optimization of the time-dependence of $\cos\theta(t)$ in Eq.(\ref{eq19}) yields the dark state: 
\begin{equation}
|D(t)|^{2} = \frac{\kappa^{2}}{\gamma}\int_{t_{0}}^{t}\Phi_{in}^{2}(t')dt', \label{eq23a}
\end{equation}
where, as already indicated, $\kappa$ is the coupling of the incoming wave-packet of the free-field into the cavity dark state, $\gamma$ is the bare cavity decay which are related as $\sqrt{c/L}=\kappa/\sqrt{\gamma}$. An interesting result of the impedance matching condition is the role it plays in optimizing the time dependence such that the dark-state amplitude tends to unity for each incoming wave-packet(s). Formula (\ref{eq23a}) in particular shows that the explicit dependence of $\cos\theta(t)$ on the shape of the input pulse $\Phi_{in}$ during the transfer process, ensures complete storage (i.e. a total transferability) of the input photon field in the cavity-dark state.

For the retrieval process, an adiabatic rotation of the mixing angle releases the stored photons into free-field photons at some later time $t_{1}$. It is relevant to point out that the resulting wave-packet will not necessarily have the same pulse form as the original one. However, the output wave-packet is generated in a well defined form and should correspond, in the ideal limit, to a single-photon Fock state. Therefore, for a time $t_{1}$ large enough such that the input wave-packet $\Phi_{in}(t)$ is "completely" stored (i.e. $\Phi_{in}(t)=0$ for $t>t_{1}$), and for $\cos\theta(t_{1})=0$, we find from the input-output relation:
\begin{equation}
\Phi_{out}(0,t) = -\sqrt{\frac{\gamma L}{c}}D(t_{1})\cos\theta(t)e^{-\frac{\gamma}{2}\int_{t_{1}}^{t}\cos^{2}\theta(\tau')d\tau'},\label{eq23b}
\end{equation}
where $D(t_{1})$ is the dark state population at the retrieval time $t_{1}$. By adiabatically switching on the Rabi frequency of the driving field, we obtain the generalized form for the output field $\Phi_{out}(t)$ for any input field i.e.:
\begin{equation}
\Phi_{out}(t) = - \frac{\Phi_{in}(t)}{\int\nolimits_{t_0}^{t}\Phi_{in}(t')dt'}\int_{t_{0}}^{t_1}\Phi_{in}^{2}(t')dt'.\label{eq25a}
\end{equation}
Eq.(26) shows a correspondance between the input and output wave-packet due to time reversal of $\cos\theta(t)$. The retrieval of the input wave-packet occurs at the time $t_{1}$ and Eq.(26) can be written in a closed form as:
\begin{equation}
\Phi_{out}(t) = - \Phi_{in}(t)\frac{|D(t_{1})|^{2}}{|D(t)|^{2}}.\label{eq25b}
\end{equation}

To check the consistency of the proposed total transfer scheme, and to point out the implications of optimizing the time-dependence of $\cos\theta(t)$ by switching off and on the Rabi frequency of the driving field in an adiabatic fashion, we consider the storage and retrieval of a Gaussian single-photon input pulse sketched in Fig.\ref{fig1}, whose normalized intensity profile is given by:
\begin{figure}\centering
\includegraphics[width=3.5in, height=2.5in]{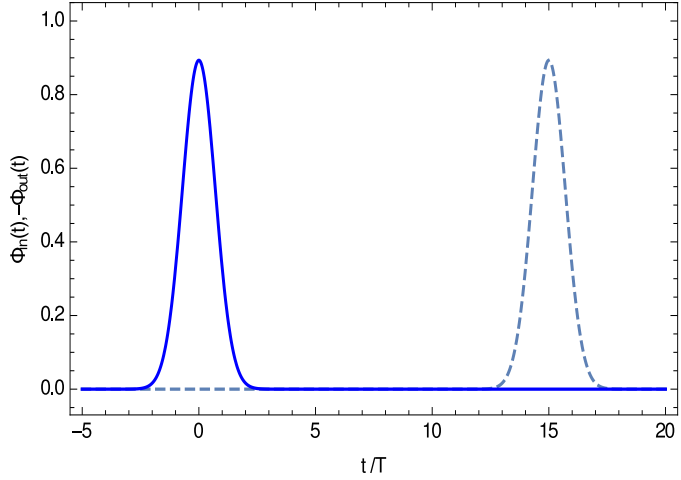}
\caption{\label{fig1} Plot of the input (full line) and output (dotted line) profiles for the Guassian pulse as a function of time. The retrieval time $t\approx 15T$.}
\end{figure}
\begin{equation}
\Phi_{1}(t)= \Phi_{in}^{(1)}(z=0,t)= \sqrt{\frac{L}{cT}}\left(\frac{2}{\pi}\right)^{1/4}exp\left[-\frac{t^{2}}{T^{2}}\right], \label{eq25c}
\end{equation}
where $T$ is the characteristic time. The storage of the input Gaussian wave packet is accomplished by changing the Rabi frequency of the driving field according to Eq.\ref{eq23}, resulting in the following optimized time-dependent $\cos\theta(t)$:
\begin{equation}
\cos\theta(t) = \sqrt{\frac{2}{\gamma T}}\left(\frac{2}{\pi}\right)^{1/4}\frac{exp\left[-\frac{t^{2}}{T^{2}}\right]}{\sqrt{1 + erf\left[\sqrt{2}\frac{t}{T}\right]}},
\label{eq25d}
\end{equation}
where $erf()$ is the Gauss error function. A consequence of changing the Rabi frequency of the driving field by varying the mixing angle $\theta(t)$, is the time evolution of the dark-state population i.e.: 
\begin{equation}
|D(t)|^{2} = \frac{1}{2}\left(1 + erf\left[\sqrt{2}\frac{t}{T} \right]\right), \label{eq25e}
\end{equation}
which approaches unity as $t\rightarrow \infty$ (see fig.\ref{fig2}). 
\begin{figure}\centering
\includegraphics[width=3.5in, height=2.5in]{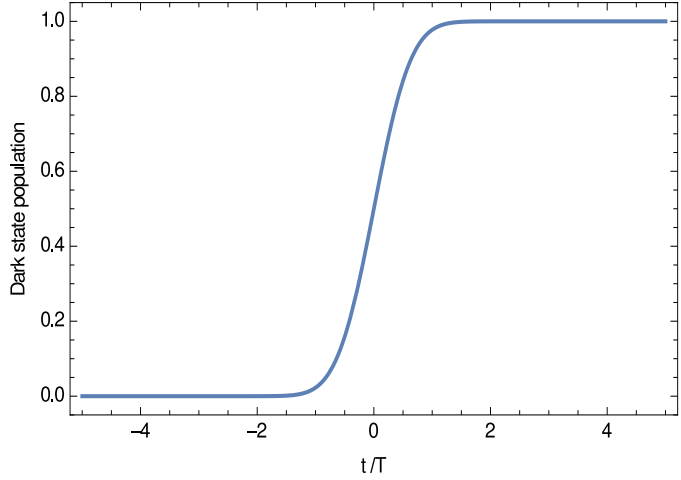}
\caption{\label{fig2} Time evolution of the dark-state population for the input Guassian profile. Population asymptotically tends to unity.}
\end{figure}

The retrieval of the stored photon into free-field photons occurs at some later time $t_{1}$. Thus, by simply reversing the adiabatic rotation of the mixing angle and using Eq.\ref{eq25a} we obtain the output field:
\begin{equation}
\Phi_{out}^{(1)}(t) = - \sqrt{\frac{L}{cT}}\left(\frac{2}{\pi}\right)^{1/4}\frac{1 + erf\left[\sqrt{2}\frac{t_{1}}{T} \right]}{1 + erf\left[\sqrt{2}\frac{t}{T} \right]}\, exp\left[-\frac{t^{2}}{T^{2}}\right].\label{eq25f}
\end{equation}
Profile of the output field $\Phi_{out}^{(1)}(t)$ is sketched in fig.\ref{fig1}, assuming a retrieval time $t\approx 15 T$.

\section{Storage and retrived of time-entangled input pulses train}
\label{secthree}
In the previous section we developed a complete formalism for the storage and subsequent retrieval of a an input photon field in the quantum states $k$ of a three-level $\Lambda$ atoms system. We obtained the analytic expressions of the dark-state population, the optimized mixing angle and the output field as explicit functnput field. We illustrated the consistency of the proposed scheme by considering a Gaussian wave packet and found that for this specific input field, the dark-state population was kink shaped in time with an asymptotic value of one as $t\rightarrow \infty$. 

In this section we use the above analytical results to probe the possibility to store and retrieve a train of single-pulse photon solitons, periodically loaded in the three-level cold atom system at a finite and constant time period. To this last point, while the Gaussian field considered in the previous section are pulse shaped they are nevertheless lower-intensity fields, unlike hyperbolic-secant (i.e. a "sech") pulses which are waves of permanent profile by virtue of their soliton features. In ref. \cite{a33}, Fleischhauer have addressed the issue of storing and retrieving single "sech" pulse described by the followiwng normalized hyperbolic secant function:
\begin{equation}
\Phi_{2}(t)= \Phi_{in}^{(2)}(z=0,t)=\sqrt{\frac{L}{cT}}\,sech\left[\frac{2t}{T}\right]. \label{inp}
\end{equation}
With this input field, the dark-state population evolves in time according to the formula:
\begin{equation}
D(t) =\sqrt{\frac{1 + tanh\left[2t/T\right]}{2}}, \label{dst1}
\end{equation}
the characteristic feature of which is its smooth-out (i.e. kink) profile extending from $0$ to $1$ . The corresponding Rabi frequency follows from formula (\ref{eq23});
\begin{equation}
\Omega(t) = g\sqrt{N} \frac{sech(\frac{2t}{T})}{\sqrt{\frac{\gamma T}{2} \left[1 + tanh(\frac{2t}{T})\right] - sech^{2}(\frac{2t}{T})}}. \label{eq24}
\end{equation}
Actually the dark-state population given in formula (\ref{dst1}), is associated with the following time-dependent optimized $\cos\theta(t)$;
\begin{equation}
\cos\theta(t) = \sqrt{\frac{2}{\gamma T}}\frac{sech(\frac{2t}{T})}{\sqrt{1 + tanh(\frac{2t}{T})}}.
\end{equation}
Let us think of a transfer scheme in which not just one single "sech" photon pulse, but a packet of identical "sech" pulses of the form (\ref{inp}) is loaded in the cold-atom system, one at a time over a well-defined finite time interval say $\tau$.  When the loading period $\tau$ is short enough the input pulse train can evolve into a time-entangled pulse multiplex with the profile of a periodic lattice of pulse solitons, so-called soliton crystal \cite{a38}. Analytically we traduce this in terms of a periodic input field for which the normalized hyperbolic secant pulse Eq.(\ref{inp}) is the fundamental component i.e.:
\begin{equation}
\Phi_{in}(t)= \sqrt{\frac{L}{cT}}\,\sum_{\ell=0}^M {sech\left[\frac{2}{T}(t-\ell\,\tau)\right]}, \label{inp1}
\end{equation}
corresponding precisely to a train of $M$ hyperbolic-secant pulses periodically loaded into the atom-cavity system at a finite time interval $\tau$. To find the corresponding Rabi frequency and dark-state population, let us assume the soliton train contains an infinite number of pulses and that the temporal separation $\tau$ between neighbour pulses is sufficiently short compared to their propagation time. With this assumption, the sum formula Eq.(\ref{inp1}) becomes exact~\cite{a38} leading to:
\begin{equation}
\Phi_{3}(t)=\Phi_{in}^{(3)}(z=0,t)=\frac{2K(m')}{\pi}\sqrt{\frac{L}{cT}}\,dn\left[\frac{4K(m')}{\pi}\,\frac{t}{T}\right], \label{eq25}
\end{equation}
where $dn$ is a Jacobi elliptic function \cite{a43} and its modulus $m$ is uniquely determined by the transcendental equation \cite{a44}:
\begin{equation}
\tau= \frac{\pi K(m)}{2 K(m')}\,T, \label{eq:25a}
\end{equation}
supplemented with the constraints $0\leq m \leq 1$, $m'=1-m$. $K(m)$ is the complete elliptic integral of the first kind \cite{a43}. 

\begin{figure}\centering
\includegraphics[width=3.5in, height=2.5in]{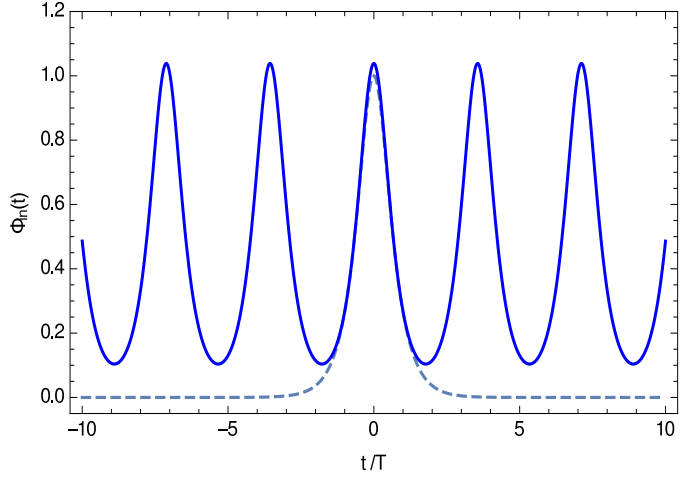}
\caption{\label{fig5} Plot of the input soliton train (\ref{eq25}) as a function of time: $m=0.99$ (full line), $m=1$ (dotted line).}
\end{figure}

\begin{figure}\centering
\includegraphics[width=3.5in, height=2.5in]{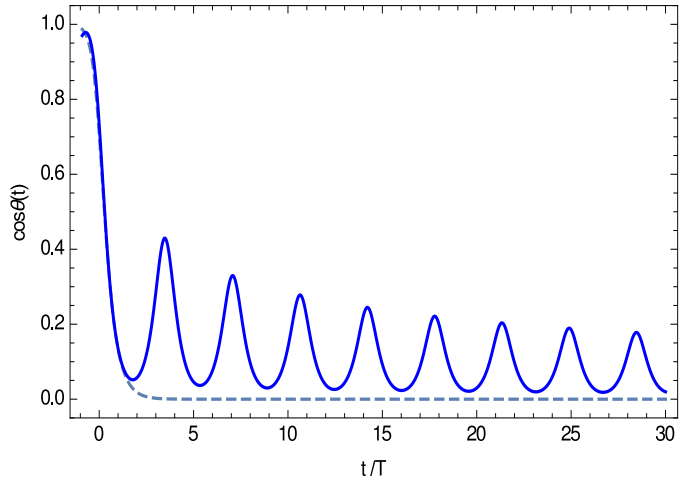}\
\caption{\label{fig6} Time evolution of $\cos\theta(t)$ given by (\ref{28}) ($\gamma T=3.8$), for $m=0.99$ (full line) and $m=1$ ($\gamma T=4$)(dotted line).}
\end{figure}

\begin{figure}\centering
\includegraphics[width=3.5in, height=2.5in]{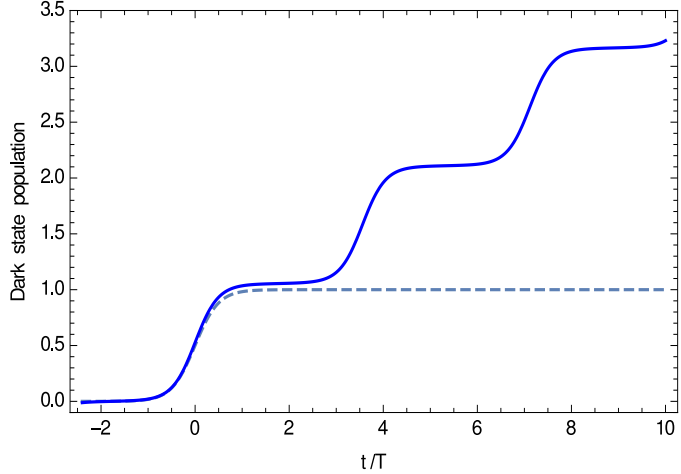}
\caption{\label{fig7} Time evolution of the dark-state population formula (\ref{eq30}), for $m=0.99$ (full line) and $m=1$ (dotted line).}
\end{figure}

Fig.\ref{fig5} represents the temporal profile of the soliton-crystal photon input field Eq.(\ref{eq25}), for $m=0.99$ (full line) and $m=1$ (dotted line).  
When $m=0$ the $dn()$ function tends to $sin()$ whereas for $m=1$, the temporal separation $\tau$ between two adjacent pulses in the train tends to infinity. In this limit Eq.(\ref{eq25}) reduces exactly to Eq.(\ref{inp}).\\
Replacing formula (\ref{eq25}) in Eqs.(\ref{eq22}) and (\ref{eq23}) we find:
\begin{equation}
\cos\theta(t) = \sqrt{\frac{\chi}{\gamma T}}\frac{dn\left[\chi \frac{t}{T}\right]}{\sqrt{E\left[am(\chi\frac{t}{T}),m\right] -E\left[am(\chi\frac{t_0}{T}),m\right]}}, \label{eq26}
\end{equation}
where $\chi = \frac{4K(m')}{\pi}$ and  $am$ is the Jacobi amplitude function, and:
\begin{equation}
\Omega(t) = \frac{g\sqrt{N}dn\left[\chi\frac{t}{T}\right]}{\sqrt{\frac{ \gamma T}{\chi}\left[E\left(am(\chi\frac{t}{T}),m\right) -E\left(am(\chi\frac{t_0}{T}),m\right)\right] - dn^{2}\left[\chi,\frac{t}{T}\right]}}, \label{eq27}
\end{equation}
corresponding respectively to the temporal evolutions of the cosine of the mixing angle $\theta(t)$ and the Rabi frequency. As these two quantities depend on the arbitrary parameter $t_0$, to gain the physics in their analytical expressions we need to fix this parameter. Relying on their asymptotic forms in the single-pulse regime i.e. when $m=1$, we choose $t_0=-\tau$ given that when $m=1$ the initial loading time $t_0\rightarrow -\infty$ consistently with the single-pulse case \cite{a33}. With this choice Eqs.(\ref{eq26}) and (\ref{eq27}) become respectively: 
\begin{equation}
\cos\theta(t) = \sqrt{\frac{\chi}{\gamma T}}\frac{dn\left[\chi\frac{t}{T}\right]}{\sqrt{E(m) + E\left[am\left(\chi\frac{t}{T}\right),m\right]}},\label{28}
\end{equation}
\begin{equation}
\Omega(t) = \frac{g\sqrt{N}dn\left[\chi\frac{t}{T}\right]}{\sqrt{\frac{ \gamma T}{\chi}\left[E(m) + E\left(am[\frac{t}{T}],m\right)\right] - dn^{2}\left[\chi\frac{t}{T}\right]}}. \label{29}
\end{equation}
In fig. \ref{fig6}, we plotted the time evolution of $\cos\theta(t)$ given by Eq.(\ref{28}) for $m=0.99$ (full line) and $m=1$ (dotted lines). Note that the dotted curve is the temporal profile of $\cos\theta(t)$ for the single-pulse input field. $\cos\theta(t)\rightarrow 0$ as $t\rightarrow\infty$, in agreement with the impedance matching condition. \\
As for the dark-state population, whose time evolution is readily expected to provide the most enlightening insight onto the proposed storage-retrieval process involving the input pulse train, this quantity is obtained by replacing Eq.(\ref{eq25}) in Eq.(\ref{eq23a}) and integrating we find:
\begin{equation}
|D(t)|^{2} = \frac{\chi}{4}\left[E(m) + E\left(am[\chi\frac{t}{T}],m\right)\right]. \label{eq30}
\end{equation}
The last quantity is plotted in fig. \ref{fig7} versus time, for $m=0.99$ (full line) and $m=1$ (dotted line). 
For $m\neq 1$, the time evolution of the dark-state population is a periodic train of kinks. The period of the kink lattice is actually the time $\tau$ separating two consecutive complete loadings of two identical single-pulse photons. As the figure indicates, over this time scale the dark-state population is fully kink shaped. As $m\rightarrow 1$ the separation between kinks $\tau \rightarrow \infty$, the dark-state population in this limit changes in time as a single kink.

The retrieval occurs at a later time $t_{1}$ when the input pulse train had been trapped in the dark state, such that no field is observed within the cavity. In this retrieval process, the action of the mixing angle is reversed by adiabatically switching on the classical driving field according to Eq.\ref{29} at some desired time. From Eq.\ref{eq25a} we find:
\begin{equation}
\Phi_{out}^{(3)}(t) =-\frac{2K(m')}{\pi} \sqrt{\frac{L}{cT}} dn\left[\chi\frac{t}{T} \right] \frac{E(m) + E\left[am\left(\chi\frac{t_1}{T}\right) ,m\right]}{ E(m) + E\left[am\left(\chi\frac{t}{T}\right) ,m\right]}.
\end{equation}
\begin{figure}\centering
\includegraphics[width=3.5in, height=2.5in]{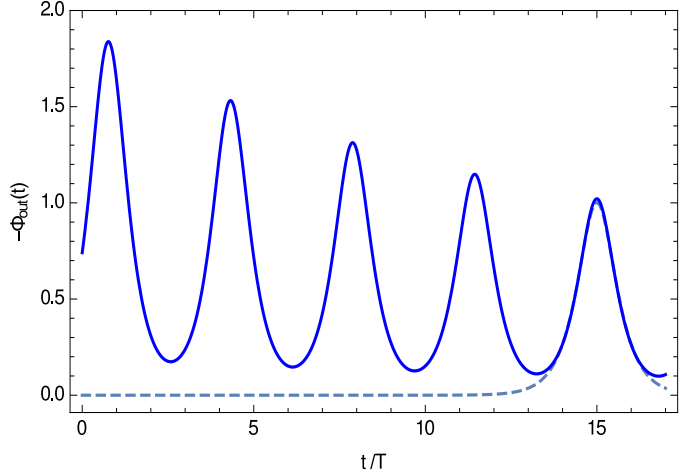}
\caption{\label{fig8} Temporal profile of the output soliton train at $t\approx 15T$ with optimized $\cos\theta(t)$, for $m=0.99$ (full line) $m=1$ (dotted line).}
\end{figure}
The output profile is plotted in fig. \ref{fig8} for $m=0.99$ (full line) and $m=1$ (dotted lines). The plot shows the retrieval at a time $t\approx 15T$ at which each stored pulse is released into free-field photons. Interestingly, the retrived pulse train is identical in shape to the input pulse train. 

\section{Conclusion}
\label{secfour}
Multi-soliton structures have attracted a great deal of attention in recent years, because of their enormous potential in information processing using multiplexed single-mode or multi-mode optical pulses. One of their many virtues is the possibility to simultaneously transmit either identical signals or several distinct signals, to different users. In this work, we investigated the storage and subsequent retrieval of a train of identical high-intensity photons in collective atomic states, by means of the adiabatic transfer mechanism. This process is based on the intracavity EIT, by which properties of a cavity filled with three-level $\Lambda$-type atoms can be manipulated by an external classical driving field. Since the mapping of the quantum information contained in the photon field into collective atomic states is achieved using the technique of intracavity EIT, it is non-destructive and reversible. By varying the mixing angle in a specific way, one can store and retrieve wave packets into the dark state of collective atomic system by switching adiabatically off and on the Rabi frequency of the classical driving field. The mechanism of adiabatic photon transfer under the quantum impedance matching condition (i.e. a one-to-one mapping of the input field into the dark state) has been discussed  in ref. \cite{a33}. In this previous work the authors applied the mechanism to the transfer of a "sech"-type pulse. We derived analytic relations for the optimization of $\cos\theta(t)$ and the corresponding Rabi frequency $\Omega(t)$, and obtained a general relation between the dark-state population and any form of input wave packet $\Phi_{in}$. These relations are valid for any given wave packet, but provided that the arrival time $t_{0}$ is well defined. 

\section*{Acknowledgments}
A. M. Dikand\'e is Alexander von Humboldt Stiftung Alumni.


\section*{References}
\begin{thebibliography}{00}
\bibitem{a1}  M. Peccianti et al., Opt. Lett. \textbf{27} (2002) 1460.
\bibitem{a2} C. Conti, M. Peccianti and G. Assanto, Phys. Rev. Lett. \textbf{92} (2004) 113902.
\bibitem{a3} C. Rotschild et al., Nature Phys. \textbf{2} (2006) 769.
\bibitem{a4} C. Conti, M. Peccianti and G. Assanto, Phys. Rev. Lett. \textbf{91} (2003) 073901.
\bibitem{a5} A. M. Dikand\'e, Europhysics Letters \textbf{94} (2011) 44004.
\bibitem{a6} M. Fleischhauer et al., Rev. Mod. Phys. \textbf{77} (2005) 633.
\bibitem{a7} L. V. Hau, S. E. Harris, Z. Dutton and C. H. Behroozi, Nature \textbf{397} (1999) 594.
\bibitem{a8} C. Liu, Z. Dutton, C. H. Behroozi and L. V. Hau, Nature \textbf{409}(6819) (2001) 490.
\bibitem{a9} M. Fleischhauer, A. Imamoglu and J. P. Marangos, Rev. Mod. Phys. \textbf{77} (2005) 633.
\bibitem{a10} M. M. Kash et al., Phys. Rev. Lett. \textbf{82} (1999) 5229.
\bibitem{a11} P. C. Ku, C. J. Chuang-Hasnain and S. L. Chuang, J. Phys. D \textbf{40} (2007) R93.
\bibitem{a12} H. Schmidt and A. Imamoglu, Opt. Lett. \textbf{21} (1996) 1936.
\bibitem{a13} S. E. Harris and L. V. Hau, Phys. Rev. Lett. \textbf{82} (1999) 4611.
\bibitem{a14} H. Wang, D. Goorskey and M. Xiao, Phys. Rev. Lett. \textbf{87} (2001) 073601.
\bibitem{a15} E. Ignesti, R. Buffa, L. Fini, E. Sali, M. V. Tognetti and S. Cavalieri, Optics Commun. \textbf{285} (2012) 1185.
\bibitem{a16} S. E. Harris, J. E. Field and A. Imamoglu, Phys. Rev. Lett. \textbf{64} (1990) 1107.
\bibitem{a17} M. D. Lukin, P. R. Hemmer, M. Loffler and M. O. Scully, Phys. Rev. Lett. \textbf{81} (1998) 2675.
\bibitem{a18} H. Li and G. Huang, Phys. Rev. A \textbf{76} (2007) 043809.
\bibitem{a19} H. A. Haus and W. S. Wong, Rev. Mod. Phys. \textbf{68} (1996) 423.
\bibitem{a20} Y. S. Kivshar and B. Luther-Davies, Phys. Rep. \textbf{298} (1998) 81.
\bibitem{a21} G. P. Agrawal, Nonlinear Fiber Optics (Academic, New York, 2001), 3rd ed.
\bibitem{a22} G. Huang, L. Deng and M. G. Payne, Phys. Rev. E. \textbf{72} (2005) 016617.
\bibitem{a23} J. Dilley, P. Nisbet-Jones, B. W. Shore and A. Kuhn, Phys. Rev. A 85 (2012) 023834.
\bibitem{a24} Y. Wu and L. Deng, Opt. Lett. \textbf{29} (2004) 2064.
\bibitem{a25} L. G. Si, W. X. Yang, X.-Y. Lü, X. Hao, and X. Yang, Phys. Rev. A \textbf{82} (2010) 013836.
\bibitem{a26} Y. Chen, Z. Bai and G. Huang, Phys. Rev. A \textbf{91} (2014) 023835.
\bibitem{a27} M. V. Gromovyi, V. I. Romanenko, S. Mieth, T. Halfmann and L. P. Yatsenko, Optics Commun.\textbf{284} (2011) 5710.
\bibitem{a28} G. Vemuri and K. V. Vasavada, Optics Commun. \textbf{129} (1996) 379.
\bibitem{a29} V. V. Kozlov and J. H. Eberly, Optics Commun. \textbf{179} (2000) 85.
\bibitem{a30} G. Buica and T. Nakajima, Optics Commun. \textbf{332} (2014) 59.
\bibitem{a31} Y. Chen, Z. Chen and G. Huang, Phys. Rev. A \textbf{89} (2015) 023820.
\bibitem{a32} M. D. Lukin, S. F. Yelin and M. Fleischhauer, Phys. Rev. Lett. \textbf{84} (2000) 4232.
\bibitem{a33} M. Fleischhauer, S. F. Yelin and M. D. Lukin, Optics. comm. \textbf{179} (2000) 395.
\bibitem{a34} M. D. Lukin, M. Fleischhauer, M. O. Scully and V. L. Velichansky, Opt. Lett. \textbf{23} (1998) 295.
\bibitem{a35} C. K. Law and H. J. Kimble, J. Mod. Opt. \textbf{44} (1997) 2067.
\bibitem{a36} A. M. Dikand\'e, Phys. Rev. A \textbf{81} (2010) 013821.
\bibitem{a37} A. M. Dikand\'e, J. Opt. \textbf{13} (2011) 035203.
\bibitem{a38} D. Jr. Fandio Jubgang, A. M. Dikand\'e and A. Sunda-Meya, Phys. Rev. A \textbf{92} (2015) 053850.
\bibitem{a39} D. Jr. Fandio Jubgang and A. M. Dikandé, J. Opt. Soc. Am. B \textbf{34} (2017) 2721.
\bibitem{a40} S. Petrosyan and Y. Malakyan, Phys. Rev. A \textbf{88} (2013) 063817.
\bibitem{a41} R. Dicke, Phys. Rev. \textbf{93} (1994) 99.
\bibitem{a42} C. Ottaviani, S. Rebi´c, D. Vitali and P. Tombesi, Eur. Phys. J. D \textbf{40} (2006) 281.
\bibitem{a43} M. Abramowitz and I. A. Stegun, "Handbook of Mathematical Functions" (Dover Publications), 1965.
\bibitem{a44}B. A. Malomed, Phys. Rev. B \textbf{38} (1988) 9242.
\bibitem{a45} C. Cohen-Tannoudji and S. Reynaud, J. Phys. B \textbf{10} (1977) 2311.

\end{thebibliography}
\end{document}